\documentclass[prd,aps,
twocolumn,superscriptaddress,preprintnumbers,nofootinbib]{revtex4-1}
\usepackage{pslatex}
\usepackage[pdftex]{graphicx}
\usepackage{psfrag}
\usepackage{epsfig}
\usepackage{color}
\usepackage{cancel}
\usepackage{slashed}
\usepackage{amssymb}
\usepackage{amsmath}
\usepackage{hyperref}
\usepackage{xspace}

\newcommand\SARAH{{\tt SARAH}\xspace}
\newcommand\SPheno{{\tt SPheno}\xspace}

\begin{document}
\preprint{CERN-PH-TH-2015-058,IFIC/15-20}

\title{Shedding light on the $b \to s$ anomalies with a dark sector}

\author{D. Aristizabal Sierra}%
\email{daristizabal@ulg.ac.be}%
\affiliation{IFPA, Dep. AGO, Universit\'e de Li\`ege, Bat B5, Sart
  Tilman B-4000 Li\`ege 1, Belgium}
\author{Florian Staub}%
\email{florian.staub@cern.ch}%
\affiliation{Theory Division, CERN 1211 Geneva 23, Switzerland}
\author{Avelino Vicente}%
\email{Avelino.Vicente@ulg.ac.be}%
\affiliation{IFPA, Dep. AGO, Universit\'e de Li\`ege, Bat B5, Sart
  Tilman B-4000 Li\`ege 1, Belgium}
\affiliation{Instituto de F\'{\i}sica Corpuscular (CSIC-Universitat de Val\`{e}ncia), Apdo. 22085, E-46071 Valencia, Spain.}

\begin{abstract}
The LHCb collaboration has recently reported on some anomalies in $b
\to s$ transitions. In addition to discrepancies with the Standard
Model (SM) predictions in some angular observables and branching
ratios, an intriguing hint for lepton universality violation was
found. Here we propose a simple model that extends the SM with a dark
sector charged under an additional $U(1)$ gauge symmetry. The
spontaneous breaking of this symmetry gives rise to a massive
$Z^\prime$ boson, which communicates the SM particles with a valid
dark matter candidate, while solving the $b \to s$ anomalies with
contributions to the relevant observables.
\end{abstract}

\maketitle
\section{Introduction}
\label{sec:intro}

Rare decays stand among the most powerful probes of physics beyond the
Standard Model (SM). Indeed, most new physics (NP) scenarios suffer
from severe constraints due to their potentially large contributions
to flavor observables. This has motivated an intense experimental
search, with special focus on observables which are strongly
suppressed in the SM.

In 2013, the LHCb collaboration reported on the measurement of several
observables in processes involving $b \to s$ transitions. A
significant tension with the SM was found in some cases. These include
angular observables in $B \to K^\ast \mu^+ \mu^-$~\cite{Aaij:2013qta},
particularly large in case of the popular $P_5^\prime$
\cite{Matias:2012xw,DescotesGenon:2012zf,Descotes-Genon:2013vna}, as
well as a decrease, with respect to the SM expectation, in several
branching ratios~\cite{Aaij:2014pli,Aaij:2013aln}. These anomalies
received immediate attention in the flavor community, and soon several
independent global fits
\cite{Descotes-Genon:2013wba,Altmannshofer:2013foa,Beaujean:2013soa,Hurth:2013ssa}
showed that the tension could be alleviated in the presence of new
physics contributions. Interestingly, in 2014 the LHCb collaboration
also found an indication of lepton universality violation in the
theoretically rather clean ratio~\cite{Aaij:2014ora}
\begin{equation}
R_K = \frac{\text{BR}(B \to K \mu^+ \mu^-)}{\text{BR}(B \to K e^+ e^-)} = 0.745^{+0.090}_{-0.074} \pm 0.036 \, ,
\end{equation}
measured in the low dilepton invariant mass regime, which departs from
the SM result $R_K^\text{SM} = 1.0003 \pm 0.0001$ by $2.6
\sigma$~\cite{Hiller:2003js}. Again, this led to some excitement in
the community, in particular after it was found that this hint is
compatible with the previous anomalies in $b \to s$ transitions: they
can be explained by the same type of NP contributions to muonic
operators
\cite{Alonso:2014csa,Hiller:2014yaa,Ghosh:2014awa,Hurth:2014vma,Altmannshofer:2014rta}. So
far, the discussion has focused on results obtained by LHCb with an
integrated luminosity of $1$ fb$^{-1}$. The latest chapter of this
story is the recent announcement of the LHCb collaboration of new
results using the full LHC Run I dataset~\cite{LHCbtalk}, with an
integrated luminosity of $3$ fb$^{-1}$, which confirm the robustness
of the LHCb data. Indeed, the new results are compatible with those
found with $1$ fb$^{-1}$, and several theorists have used the new data
to update their analyses.  According to \cite{Straubtalk}, the
hypothesis of non-zero NP contributions is preferred over the SM by
$3.7 \sigma$, $4.3 \sigma$ if the 2014 measurement of $R_K$ is
included, whereas the analysis of \cite{Matiastalk} finds a slightly
larger statistical significance for NP (slightly larger than $4
\sigma$) even in the absence of $R_K$. In any case, assuming that
hadronic effects \cite{Jager:2012uw,Jager:2014rwa} are not behind
these anomalies (something impossible in case of $R_K$), there is
clear evidence of new physics in $B$ meson decays. For a complete
review of the subject see \cite{Altmannshofer:2014rta} and references
therein.

Several approaches have been considered in order to explain the $b \to
s$ anomalies. Some papers
\cite{Gauld:2013qba,Buras:2013qja,Buras:2014fpa} consider generic
$Z^\prime$ bosons with flavor violating couplings. These have been
shown to provide a simple way to reconcile theory predictions with
experimental data. There are also some works that were built to
address the first anomalies but fail to address the lepton
universality violating $R_K$ measurement, see for example
Refs. \cite{Gauld:2013qja,Buras:2013dea} in the context of 331
scenarios or Ref. \cite{Datta:2013kja}, where 4-quark scalar
interactions were considered. Finally, a few recent models can also
account for the lepton universality violating $R_K$ measurement. One
finds three types of working models: models with scalar or vector
leptoquarks (or similarly, R-parity violating supersymmetry, where the
squarks can play the role of the
leptoquarks)~\cite{Hiller:2014yaa,Biswas:2014gga,Gripaios:2014tna,Varzielas:2015iva},
composite Higgs models~\cite{Buras:2014fpa,Niehoff:2015bfa} and models
with a $Z^\prime$. The latter is generally considered the easiest
approach, since the anomalies can be solved with new contributions to
vectorial operators. However, to the best of our knowledge, the only
(complete) models that have been put forward in order to account for
these anomalies using a $Z^\prime$ are: (1) the one introduced in
Ref. \cite{Altmannshofer:2014cfa} (and the Two-Higgs-Doublet version
in \cite{Crivellin:2015mga}), which makes use of a $U(1)_{L_\mu -
  L_\tau}$ gauge symmetry, and (2) the one proposed in the recent
paper \cite{Crivellin:2015lwa}, which extends the horizontal gauge
symmetry to the quark sector.

The existence of dark matter (DM) is a well established evidence of
new physics, supported by a plethora of astrophysical and cosmological
observations. This motivates further extensions of the SM of particle
physics that include valid DM candidates. In DM models, the current
Planck $3 \sigma$ limits for the DM relic density, $0.1118 <
\Omega_{\text{DM}} h^2 < 0.1280$ \cite{Ade:2013zuv}, typically imply
strong constraints on the mass and couplings of the DM
candidate. This, combined with additional constraints (such as those
from flavor physics), sometimes leads to very predictive scenarios
where the parameter space of the DM model shrinks to small regions
compatible with all observations.

We propose a simple model which captures the main ingredients required
to explain the $b \to s$ anomalies and, simultaneously, provides a DM
candidate. In order to do so, we extend the SM with a dark sector
charged under an additional $U(1)$ gauge symmetry. The spontaneous
breaking of this symmetry gives rise to a massive $Z^\prime$ boson,
which communicates the SM particles with the dark matter particle and
solves the $b \to s$ anomalies with contributions to the relevant
flavor observables. The interplay between DM and flavor physics leads
to a very constrained scenario. In particular, achieving the measured
DM relic density while providing the required NP contributions to $B$
meson decays turns out to be very restrictive on the model parameters.

The rest of the paper is organized as follows: in
Sec. \ref{sec:general} we discuss some general aspects about the $b
\to s$ anomalies and introduce the basic language to be used
throughout the paper. In Sec. \ref{sec:model} we define our model,
show how it addresses the $b \to s$ anomalies while providing a DM
candidate and discuss extensions to generate non-zero neutrino
masses. Sec. \ref{sec:dm} is devoted to the DM phenomenology of our
model and the $Z^\prime$ portal responsible for its production in the
early universe. Sec. \ref{sec:constraints} is a review of the most
relevant constraints in our model, to be considered in
Sec. \ref{sec:results}, where we present our numerical
results. Finally, in Sec. \ref{sec:conclusions} we summarize our
results, discuss some related aspects and derive some general
conclusions.

\section{The $\boldsymbol{b \to s}$ anomalies}
\label{sec:general}

In this Section we discuss some general aspects about the $b \to s$
anomalies. For a complete review of the subject, we refer the reader
to \cite{Altmannshofer:2014rta}.

\subsection{Operators and global fits}
\label{subsec:operators}

The effective Hamiltonian for $b \to s$ transitions is usually written
as
\begin{equation}
\mathcal H_{\text{eff}} = - \frac{4 G_F}{\sqrt{2}} \, V_{tb} V_{ts}^\ast \, \frac{e^2}{16 \pi^2} \, \sum_i \left(C_i \mathcal O_i + C^\prime_i \mathcal O^\prime_i \right) + \text{h.c.} \, ,
\end{equation}
where $G_F$ is the Fermi constant, $e$ the electric charge and $V$ the
CKM matrix. $\mathcal O_i$ and $\mathcal O^\prime_i$ are the effective
operators that contribute to $b \to s$ transitions, and $C_i$ and
$C^\prime_i$ their Wilson coefficients. Since the most important
anomalies have been found in semileptonic $B$ meson decays, we will
consider the following set of operators,
\begin{align}
\mathcal O_9 &= \left( \bar s \gamma_\mu P_L b \right) \, \left( \bar \ell \gamma^\mu \ell \right) \, ,   &   \mathcal O^\prime_9 &= \left( \bar s \gamma_\mu P_R b \right) \, \left( \bar \ell \gamma^\mu \ell \right) \, , \label{eq:O9} \\
\mathcal O_{10} &= \left( \bar s \gamma_\mu P_L b \right) \, \left( \bar \ell \gamma^\mu \gamma_5 \ell \right) \, ,   &   \mathcal O^\prime_{10} &= \left( \bar s \gamma_\mu P_R b \right) \, \left( \bar \ell \gamma^\mu \gamma_5 \ell \right) \, . \label{eq:O10}
\end{align}
Here $\ell = e, \mu, \tau$~\footnote{When referring to operators
  involving a particular lepton flavor, we will denote it with a
  superscript, e.g. $C_9^\mu$ and $\mathcal O_9^\mu$, for
  muons.}. It is also customary to split the Wilson coefficients in
two parts: the SM contributions and the NP contributions. Since the
primed operators, $\mathcal O^\prime_9$ and $\mathcal O^\prime_{10}$,
do not receive significant SM contributions, this is usually applied
only to the unprimed Wilson coefficients, which can be written as
\begin{eqnarray}
C_9 &=& C_9^{\text{SM}} + C_9^{\text{NP}} \, , \\
C_{10} &=& C_{10}^{\text{SM}} + C_{10}^{\text{NP}} \, .
\end{eqnarray}
The SM contributions, $C_9^{\text{SM}}$ and $C_{10}^{\text{SM}}$, have
been computed by different groups. Assuming that these are the only
contributions to the Wilson coefficients, several independent global
fits have found a sizable tension with experimental data on $b \to s$
transitions. This motivates the addition of NP contributions. When
this is done, the global fits are clearly improved. According to
\cite{Altmannshofer:2014rta} (we will mainly consider the results of
this global fit), the best improvements are found in two cases:

\begin{itemize}
\item {\bf Scenario 1:} NP provides a negative contribution to $\mathcal
  O_9^\mu$, with $C_9^{\mu , \text{NP}} \sim - 30 \% \times
  C_9^{\mu , \text{SM}}$, leading to a Wilson coefficient $C_9^\mu$
  significantly smaller than the one in the SM.
\item {\bf Scenario 2:} NP enters in the $SU(2)_L$ invariant direction
  $C_9^{\mu , \text{NP}} = - C_{10}^{\mu , \text{NP}}$, with $C_9^{\mu
  , \text{NP}} \sim - 12 \% \times C_9^{\mu , \text{SM}}$.
\end{itemize}

In both cases, the rest of operators involving muons are perfectly
compatible with the SM expectations. Similarly, no NP is required for
operators involving electrons or tau leptons.

\subsection{Model building requirements}
\label{subsec:modelbuilding}

Once determined the type of contributions to $C_9$ and $C_{10}$ a NP
model has to induce, one can figure out the main ingredients of a
simple working model. Arguably, the simplest one contains the
following elements:

\begin{itemize}
\item A massive $Z^\prime$ boson, responsible for the vectorial
  operators $\mathcal O_9$ and $\mathcal O_{10}$
\item The $Z^\prime$ must have flavor violating couplings to quarks
\item The $Z^\prime$ must couple differently to electrons and muons
\end{itemize}

This setup can be easily parameterized by the Lagrangian
\cite{Buras:2012jb,Altmannshofer:2014rta}
\begin{equation}
\mathcal L \supset \bar f_i \gamma^\mu \left( \Delta_L^{f_i f_j} P_L + \Delta_R^{f_i f_j} P_R \right) f_j Z_\mu^\prime \, .
\end{equation}
In order to account for the anomalies, one requires $\Delta_L^{bs}
\neq 0$ and either (1) $\Delta_L^{\mu \mu} = \Delta_R^{\mu \mu} \neq
0$, or (2) $\Delta_L^{\mu \mu} \neq 0$ and $\Delta_R^{\mu \mu} = 0$,
depending on the scenario one wants to consider. The rest of the
$Z^\prime$ couplings to SM fermions can be set to zero. As we will see
in Sec. \ref{sec:model}, the model we are going to consider belongs to
scenario (2). In this case, the quark and lepton currents contributing
to $\mathcal O_9$ and $\mathcal O_{10}$ are both left-handed and one
finds at tree-level~\cite{Altmannshofer:2014rta}
\begin{equation} \label{eq:c9c10}
C_9^{\mu , \text{NP}} = - C_{10}^{\mu , \text{NP}} = - \frac{\Delta_L^{bs} \Delta_L^{\mu \mu}}{V_{tb} V_{ts}^\ast} \, \left( \frac{\Lambda_v}{m_{Z^\prime}} \right)^2 \, ,
\end{equation}
with
\begin{equation}
\Lambda_v = \left( \frac{\pi}{\sqrt{2} G_F \alpha} \right)^{1/2}
\simeq 4.94 \, \text{TeV} \, ,
\end{equation}
where $\alpha = \frac{e^2}{4 \pi^2}$ is the fine structure
constant. Note that $\Lambda_v$ and the CKM elements appear in
Eq. \eqref{eq:c9c10} in order to normalize the Wilson coefficients as
defined in Eqs. \eqref{eq:O9} and \eqref{eq:O10}. \\

We now introduce a complete renormalizable model with these
properties.

\section{The model}
\label{sec:model}

We extend the SM gauge group with a new dark $U(1)_X$ factor, under which
all the SM particles are assumed to be singlets. The only particles
charged under the $U(1)_X$ group are the following vector-like
fermions,
\begin{align}
Q_L &= \left( {\bf 3} , {\bf 2} , \frac{1}{6} , 2 \right) \, ,  &   Q_R &= \left( {\bf 3} , {\bf 2} , \frac{1}{6} , 2 \right) \, , \\
L_L &= \left( {\bf 1} , {\bf 2} , -\frac{1}{2} , 2 \right) \, ,  &   L_R &= \left( {\bf 1} , {\bf 2} , -\frac{1}{2} , 2 \right) \, ,
\end{align}
as well as the complex scalar fields
\begin{equation}
\phi = \left( {\bf 1} , {\bf 1} , 0 , 2 \right) \, ,  \qquad \chi = \left( {\bf 1} , {\bf 1} , 0 , -1 \right) \, ,
\end{equation}
where we denote the gauge charges under $SU(3)_c \otimes SU(2)_L
\otimes U(1)_Y \otimes U(1)_X$ and the $SU(2)_L$ doublets can be
decomposed as $Q_{L,R} = \left( U , D \right)_{L,R}$ and $L_{L,R} =
\left( N , E \right)_{L,R}$.

Besides canonical kinetic terms, the new vector-like fermions have
Dirac mass terms,
\begin{equation} \label{eq:VectorMass}
\mathcal L_m = m_Q \overline Q Q + m_L \overline L L \, ,
\end{equation}
as well as Yukawa couplings with the SM fermions
\begin{equation} \label{eq:VectorYukawa}
\mathcal L_Y = \lambda_Q \overline{Q_R} \phi q_L + \lambda_L \overline{L_R} \phi \ell_L + \text{h.c.} \, ,
\end{equation}
where $\lambda_Q$ and $\lambda_L$ are $3$ component vectors. The
scalar potential takes the form
\begin{equation}
\mathcal V = \mathcal V_{\text{SM}} + \mathcal V\left( H , \phi , \chi \right) + \mathcal V\left( \phi , \chi \right) \, .
\end{equation}
Here $H$ is the SM Higgs doublet and $\mathcal V_{\text{SM}} = m_H^2
|H|^2 + \frac{\lambda}{2} |H|^4$ is the SM scalar potential. The
pieces involving the $U(1)_X$ charged scalars are
\begin{equation}
\mathcal V\left( H , \phi , \chi \right) = \lambda_{H \phi} \, |H|^2 |\phi|^2 + \lambda_{H \chi} \, |H|^2 |\chi|^2
\end{equation}
and
\begin{eqnarray}
\mathcal V\left( \phi , \chi \right) &=& m_\phi^2 |\phi|^2 + m_\chi^2 |\chi|^2 + \frac{\lambda_\phi}{2} |\phi|^4 + \frac{\lambda_\chi}{2} |\chi|^4 \nonumber \\
&& + \lambda_{\phi \chi} \, |\phi|^2 |\chi|^2 + \left( \mu \, \phi \chi^2 + \text{h.c.} \right) \, .
\end{eqnarray}
We will assume that the scalar potential is such that only the
standard Higgs boson and the $\phi$ field acquire non-zero vacuum
expectation values (VEVs),
\begin{equation}
\langle H^0 \rangle = \frac{v}{\sqrt{2}} \, , \qquad \langle \phi \rangle = \frac{v_\phi}{\sqrt{2}} \, .
\end{equation}
Therefore, the $\phi$ field will be responsible for the spontaneous
breaking of $U(1)_X$, giving a mass to the $Z^\prime$, $m_{Z^\prime} =
2 g_X v_\phi$, where $g_X$ is the $U(1)_X$ gauge coupling, and
inducing mixings between the vector-like fermions and their SM
counterparts thanks to the Yukawa interactions in
Eq. \eqref{eq:VectorYukawa}. Furthermore, after spontaneous symmetry
breaking, the resulting Lagrangian contains a remnant $\mathbb{Z}_2$
symmetry, under which $\chi$ is odd and all the other fields are
even. Therefore, $\chi$ is a stable neutral scalar, and thus a
potentially valid DM candidate. It is worth noting that the mechanism
to stabilize the DM particle does not introduce additional ad-hoc
symmetries, but simply makes use of the same $U(1)_X$ symmetry that is
required in order to give an explanation to the LHCb
observations. This goal has been achieved by breaking the continuous
$U(1)_X$ symmetry to a remnant $\mathbb{Z}_2$, something that can be
easily accomplished with a proper choice of $U(1)_X$
charges~\cite{Krauss:1988zc,Petersen:2009ip,Sierra:2014kua}.

Before concluding this section we must comment on $U(1)$ mixing. It is
well known that nothing prevents $U(1)$ factors from mixing. In the
model under consideration, this would be given by the Lagrangian term
\cite{Holdom:1985ag}
\begin{equation}
\mathcal L \supset \varepsilon \, F_{\mu \nu}^Y F^{\mu \nu}_X \, ,
\end{equation}
where $F_{\mu \nu}^{X,Y}$ are the usual field strength tensors for the
$U(1)_{X,Y}$ groups. In the presence of a non-zero $\varepsilon$
coupling, kinetic mixing between the $U(1)_X$ and $U(1)_Y$ gauge
bosons is induced. As a consequence of this, the physical $Z^\prime$
boson would couple to all particles that carry hypercharge, this is,
to all the SM fermions. This would lead to phenomenological problems
since couplings to the first generation are strongly
constrained. Therefore, we will assume that the tree-level
$\varepsilon$ coupling vanishes. This is easily justified in our model
because this term is not induced via renormalization group running if
it is zero at some high-energy scale (where one may speculate about a
ultraviolet completion).  The reason is our choice of the $U(1)_X$
charges~\cite{Fonseca:2013bua}.  In addition, we must keep the 1-loop
induced $\varepsilon$ coupling, generated in loops including heavy
vector-like quarks and leptons, under control. We find
\begin{equation}
\varepsilon_{\text{1-loop}}
\propto \frac{g_1 g_X}{16 \pi^2} \log \left( \frac{m_Q}{m_L} \right) \, .
\end{equation}
Therefore, $m_Q \sim m_L$ would ensure small $\varepsilon$ couplings,
while allowing large $g_X$. Something required by DM constraints (see
Sec. \ref{sec:dm}).

\subsection{Solving the $b \to s$ anomalies}
\label{subsec:solving}

\begin{figure}[t!]
\centering
\includegraphics[scale=0.5]{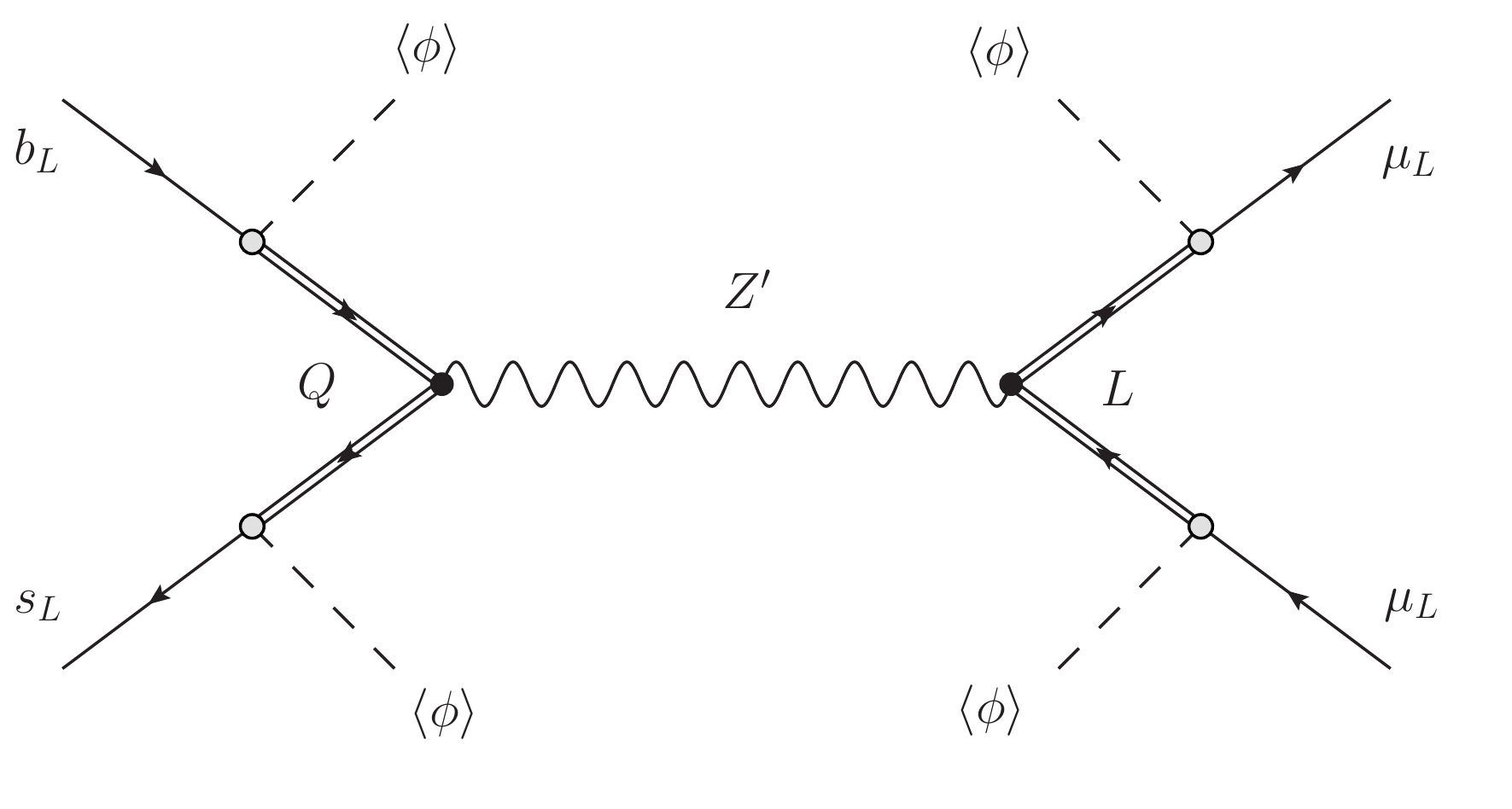}
\caption{Generation of $\mathcal O_9$ and $\mathcal O_{10}$ in our model.}
\label{fig:couplings}
\end{figure}

This model solves the $b \to s$ anomalies in a similar fashion as the
one in Ref. \cite{Altmannshofer:2014cfa}. The $Z^\prime$ couplings to
the SM fermions are generated after their mixing with the
corresponding vector-like quarks and leptons, as shown in
Fig. \ref{fig:couplings}. Neglecting $m_{s,b}^2 \ll m_Q^2$ and
$m_\mu^2 \ll m_L^2$, the resulting $Z^\prime$ couplings are found to
be
\begin{equation}
\label{eq:Delta-parameters}
\Delta_L^{bs} = \frac{2 \, g_X  \lambda_Q^{b} \lambda_Q^{s \ast} v_\phi^2}{2 m_Q^2 + \left(|\lambda_Q^{s}|^2+|\lambda_Q^{b}|^2\right) v_\phi^2} \quad , \quad
\Delta_L^{\mu \mu} = \frac{2 \, g_X |\lambda_L^{\mu}|^2 v_\phi^2}{2 m_L^2 + |\lambda_L^{\mu}|^2 v_\phi^2} \, .
\end{equation}
In case $\lambda_Q^{s,b} \ll 1$, the $Z^\prime$ coupling to a pair of
SM quarks can be further approximated to $\Delta_L^{bs} \simeq g_X
v_\phi^2 \, \frac{\lambda_Q^{b} \lambda_Q^{s \ast}}{m_Q^2}$. These
results can be combined with Eq. \eqref{eq:c9c10} in order to
determine the allowed ranges for the model parameters that explain the
$b \to s$ anomalies found by LHCb.

\subsection{Neutrino masses}
\label{subsec:numass}

\begin{figure}[t!]
\centering
\includegraphics[scale=0.5]{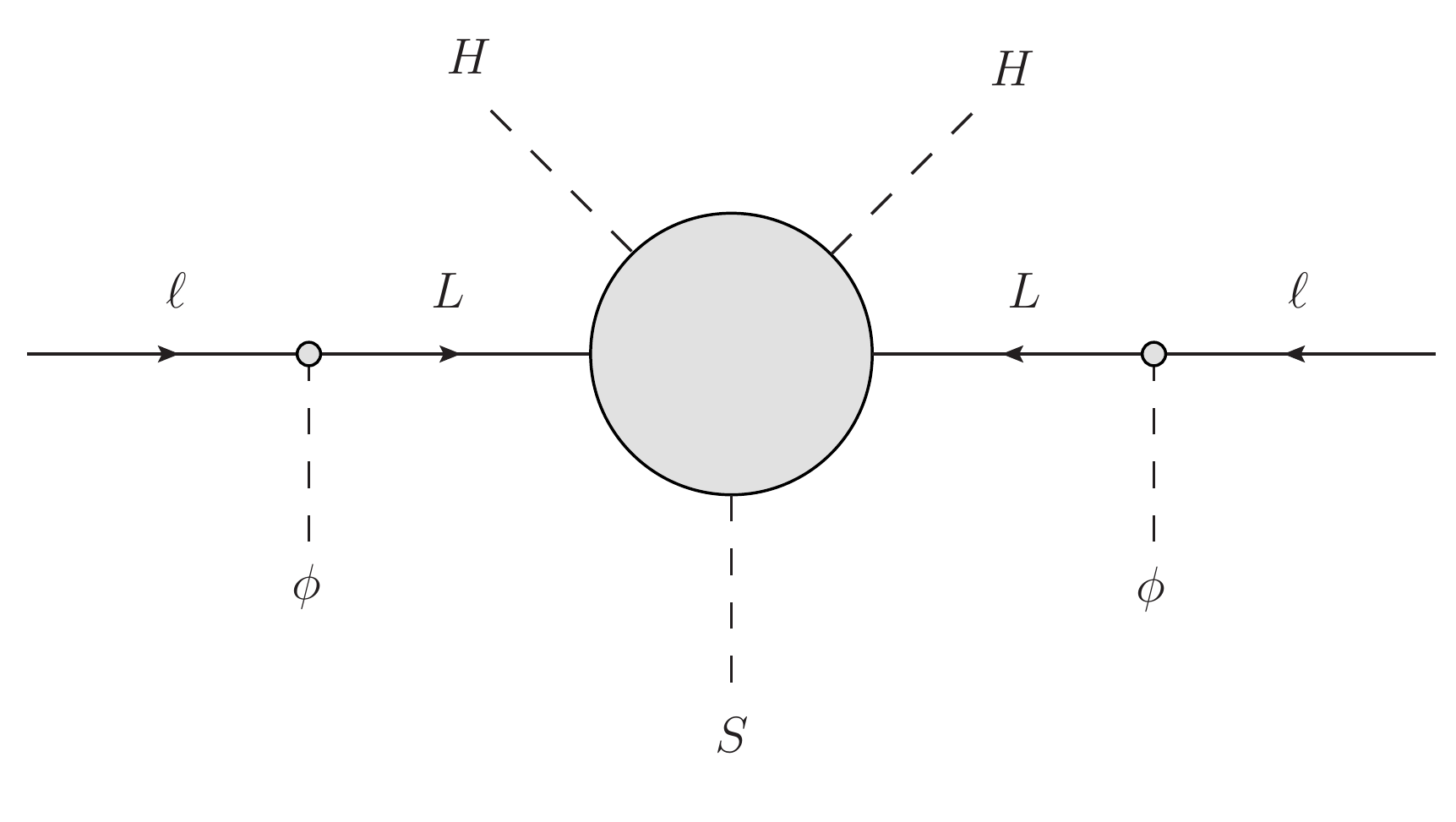}
\caption{Non-trivial neutrino mass generation in our setup. $S$ is a
  new scalar field with $U(1)_X$ charge $-4$, necessary in order to
  make the operator gauge invariant.}
\label{fig:numass}
\end{figure}

We can further extend the model to get non-zero masses for the SM
neutrinos. This can be done trivially by adding new particles,
singlets under $U(1)_X$, which mediate the standard mechanisms. For
example, the addition of right-handed neutrino singlets, $\nu_R =
\left( 1 , 1 , 0 , 0 \right)$, allows for the usual type-I seesaw
mechanism.

It is, however, more interesting to consider a mechanism that involves
the $U(1)_X$ sector of the model. This can be done by means of the
effective operator
\begin{equation}
\mathcal O_\nu = \frac{1}{\Lambda_\nu^5} \ell \ell H H \phi \phi S \, , 
\end{equation}
as shown in Fig. \ref{fig:numass}. Here $S$ is a new scalar field with
with $U(1)_X$ charge $q_S = -4$, necessary in order to make the
operator gauge invariant. An example model that can serve as
ultraviolet completion of $\mathcal O_\nu$ is obtained with the
addition of the scalar $S = \left( 1 , 1 , 0 , -4 \right)$, together
with a vector-like (Dirac) fermion $F = \left( 1 , 1 , 0 , 2
\right)$. This allows for the Yukawa couplings $\lambda_S S
\overline{F^c} F$ and $y \overline L H F$, which lead to $\mathcal
O_\nu$ after integrating out $F$ and $L$. Although $S$ must get a
non-zero VEV in order to break lepton number and generate neutrino
masses, we note that our choice $q_S = -4$ guarantees that the remnant
$\mathbb{Z}_2$ symmetry that stabilizes the DM particle $\chi$ is
preserved.

\section{Dark matter phenomenology}
\label{sec:dm}
At high temperatures $\chi$ attains thermal equilibrium in the heat
bath through its reactions with the different degrees of freedom to
which it couples. At decoupling, the dominant processes determining
the DM yield are the following $2\leftrightarrow 2$ reactions
\footnote{Note that since $m_{Z^\prime}>m_\chi$, $Z^\prime\to
  \chi\chi^*$ plays a subdominant role.}:
\begin{itemize}
\item $HH^\dagger \leftrightarrow \chi\chi^*$ (Higgs portal):
  processes enabled by the $\lambda_{H \chi} \, |H|^2 |\chi|^2$
  coupling.
\item $\bar F F\leftrightarrow \chi\chi^*$, with $F$ standing for the
  SM and the new vector-like quarks and leptons ($Z^\prime$ portal):
  the former enabled by $F_\text{SM}-F$ mixing and $U(1)_X$ coupling,
  while the latter solely by $g_X$.
\end{itemize}
For very heavy DM with $m_\chi > m_{Z^\prime}$, also $Z^\prime
Z^\prime \leftrightarrow \chi\chi^*$ can take place.  Depending on the
relative size of these processes (pure scalar and $Z^\prime$-mediated)
one can then distinguish several scenarios. Among them those solely
involving $Z^\prime$-mediated processes, are---arguably---expected to
be dominant (they are driven by a gauge coupling).  Interestingly
enough, in that case a clear correlation with flavor physics must
exist. Note that these processes match those in
Fig. \ref{fig:couplings} if one trades one of the fermion pairs for
$\chi\chi^*$. Therefore, under the fairly reasonable assumption that
the $Z^\prime$-mediated processes play a dominant role, an interplay
between flavor and DM physics is possible establishing. Thus, in turn,
further constraining the flavor-transition parameters through the
restrictions imposed by the condition of generating the correct DM
relic density (see the useful
Ref. \cite{Chu:2013jja})\footnote{Further constraints such as
  direct/indirect detection and collider searches might be also
  relevant, see \cite{Alves:2013tqa,Alves:2015pea}.}.

Although our results rely on {\tt MicrOmegas} \cite{Belanger:2013oya}, a
simple analytical discussion is worth doing. The cross section for the
$\bar F F\leftrightarrow \chi\chi^*$ processes can be estimated to be
\begin{equation}
  \label{eq:x-sec}
  \sigma(s)\sim \left|\Delta_L^{f_if_j}\right|^2g_X^2\frac{1}{s}
  \frac{m_{Z^\prime}^4}
  {\left(m_{Z^\prime}^2-s\right)^2 - m_{Z^\prime}^2\Gamma_{Z^\prime}^2}
  \,f(x_f,x_s)\ ,
\end{equation}
where the $x_i=m_i^2/s$ and $f(x_f,x_s)$ is a kinematic
function. Depending on the relative size of $m_\chi$ and $m_{Q,L}$
($r=m_\chi/m_{Q,L}$), one can distinguish two regimes. For $r<1$, DM
annihilation processes involve dominantly SM quarks and leptons ($b_L,
s_L$ and $\mu_L$). In that case, however, the corresponding cross
sections are suppressed by chiral/vector-like mixing (see
eq. (\ref{eq:Delta-parameters})), annihilation is rather inefficient
and therefore leads to an overpopulation of $\chi$ scalars. For $r>1$,
processes involving the vector-like fields are also possible, and
since they are not suppressed by mixing parameters they can lead to
the appropriate DM relic abundance. Note that if the DM and
vector-like fermion mass splitting is small, then the cross section
will be phase space suppressed: $f\to 0$. In this case, resonant
annihilation is needed to efficiently deplete the scalar $\chi$
population (see Fig. \ref{fig:C9DM}). For large mass splittings, in
contrast, annihilation is very efficient, and so the correct relic
density can be readily obtained. However, reconciling this
``scenario'' with flavor constraints turns out to be tricky. Avoiding
the resonance requires $m_{Z'}>m_\chi$, which means $m_{Z'}\gg
m_{L,Q}$, and a large $g_X$ close to 1. In that limit,
\begin{equation}
  \label{eq:C9_NP-limit}
  C_9^\text{NP}\sim -\frac{\lambda_Q^b\lambda^s_Q}{m_Q^2}\ ,
\end{equation}
which for typical vector-like quark masses, $m_Q \sim$~TeV, implies
large chiral/vector-like quark mixing (order one $\lambda_Q^{b,s}$
couplings) in order to be compatible with the LHCb observations. This,
however, is forbidden by quark flavor constraints (see next section).

In summary, the correct DM relic density can be easily produced within
our setup provided $r\sim {\cal O}(1)$ and DM annihilation proceeds
resonantly. Resonant annihilation can be avoided for $r\gg 1$, but
finding spots in parameter space consistent with both, DM and flavor
constraints seems challenging. The right relic density might as well
be produced when $r<1$, but probably this would require being sharply
at the $Z^\prime$ resonance.

\section{Constraints}
\label{sec:constraints}

There are constraints on the mass of additional gauge bosons coupled
to SM states. The ballpark of these limits is about 2.5--3.0~TeV for
$U(1)$ extensions which predict a coupling of the $Z'$ to light quarks
of $O(1)$ \cite{Accomando:2015cfa}. However, in our model these
couplings are suppressed by the small mixing between the SM quarks and
the vector-like states. Thus, we nearly get any constraint on the mass
of the $Z'$ from LHC searches. In addition, we are going to assume in
the lepton sector only a sizable mixing for muons with the new
states. Therefore, we get also hardly any limit on $m_{Z^\prime}$ from
LEP searches.

Thus, we can safely assume in the following that the $Z^\prime$
couplings to the first generation of SM fermions can be neglected and
evade all LEP and LHC limits. Similarly, $Z^\prime$ contributions to
flavor observables related to the first generation can be safely
neglected. Let us now review several relevant constraints in our
model.

\subsection{Collider constraints}

The masses of the vector-like quarks have strong bounds from the
LHC. Being colored particles, vector-like quarks can be efficiently
produced in $pp$ collisions, which typically pushes their masses
towards the TeV scale, see \cite{Barducci:2014ila} and references
therein. Since our setup works with vector-like quarks with masses at
or above the TeV scale, these bounds are easily satisfied.

Vector-like leptons can be searched for at the LHC in the standard
multi-lepton channels. Using the CMS analysis in
\cite{CMS-PAS-SUS-13-002}, based on searches for final states
including $3$ or more leptons with an integrated luminosity of about
$19.5$~fb$^{-1}$ at $\sqrt{s} = 8$~TeV, Ref. \cite{Falkowski:2013jya}
obtained a lower limit on the mass of doublet vector-like leptons of
about $460$~GeV, in case they decay to electrons or muons, and about
$280$~GeV, in case they decay to tau leptons. More recently,
Ref. \cite{Dermisek:2014qca} considered the analogous ATLAS
multi-lepton search \cite{ATLAS-multilepton}, with $20.3$~fb$^{-1}$ at
$\sqrt{s} = 8$~TeV, also looking for final states with $3$ or more
leptons. The results were similar to the ones derived from the CMS
analysis, with lower limits on the mass of doublet vector-like leptons
of about $\sim 500$~GeV.

Finally, one can also derive limits from $Z \to 4 \ell$ searches at
the LHC. However, according to \cite{Altmannshofer:2014cfa}, these are
rather mild, only relevant for $Z^\prime$ masses below $100$~GeV.

\subsection{Quark flavor constraints}

Contrary to the model in \cite{Altmannshofer:2014cfa}, where the
muonic current contributing to the $b \to s$ observables is purely
vectorial ($\bar \ell \gamma_\alpha \ell$), in our model the current
is left-handed and thus contains an axial vector contribution as well
($\bar \ell \gamma_\alpha \gamma_5 \ell$). This is relevant for the
$B_s \to \mu^+ \mu^-$ decay, especially sensitive to axial-vector
currents. Currently, there is a little tension between the SM
prediction for the average time-integrated branching ratio,
$\overline{\text{BR}}(B_s \to \mu^+ \mu^-)$, and the CMS and LHCb
measurements. While the SM prediction is found to be
$\overline{\text{BR}}(B_s \to \mu^+ \mu^-)_{\text{SM}} = (3.65 \pm
0.23) \times 10^{-9}$~\cite{Bobeth:2013uxa}, the combination of the
CMS and LHCb measurements leads to $\overline{\text{BR}}(B_s \to \mu^+
\mu^-)_{\text{exp}} = (2.9 \pm 0.7) \times
10^{-9}$~\cite{Bsmumuexp}. In view of this little deficit, the new
$Z^\prime$ mediated contribution might be potentially welcome.

The $B_s \to \mu^+ \mu^-$ decay is known for its reach in probing new
physics scenarios (see for example the recent papers
\cite{DeBruyn:2012wk,Buras:2013uqa,Yeghiyan:2013upa,Becirevic:2012fy,Dreiner:2013jta}). In
the SM, the amplitude is dominated by the axial vector leptonic
current, and thus by the $C^\mu_{10}$ Wilson coefficient ($C_{10}^{\mu
  , \text{SM}}$). Similarly, in the model under consideration, the
main NP contribution is given by a modification of $C^\mu_{10}$. We
note that vector contributions vanish for lepton flavor conserving
channels. Therefore, our prediction for the average time-integrated
branching ratio is simply given by
\begin{equation}
\overline{\text{BR}}(B_s \to \mu^+ \mu^-) = \left| \frac{C_{10}^{\mu , \text{SM}} + C_{10}^{\mu , \text{NP}}}{C_{10}^{\mu , \text{SM}}} \right|^2 \overline{\text{BR}}(B_s \to \mu^+ \mu^-)_{\text{SM}} \, .
\end{equation}
Nevertheless, the large experimental error in
$\overline{\text{BR}}(B_s \to \mu^+ \mu^-)$ precludes from obtaining
useful bounds. Using the SM and the experimental average
time-integrated branching ratios given above, one finds $-0.25 <
C_{10}^{\mu , \text{NP}}/C_{10}^{\mu , \text{SM}} < 0.03$ (at the $1
\sigma$ level). By combining $C_{10}^{\mu , \text{NP}} = - C_9^{\mu ,
  \text{NP}}$ with the SM values for $C_9^\mu$ and $C_{10}^\mu$, one
finds that in our setup $C_{10}^{\mu , \text{NP}}/C_{10}^{\mu ,
  \text{SM}} \sim 0.06$, which is slightly above the $1 \sigma$ limit,
and thus perfectly compatible with the experimental measurement of
$\overline{\text{BR}}(B_s \to \mu^+ \mu^-)$ at the $2 \sigma$ level.

Another relevant flavor constraint comes from $B_s-\bar B_s$ mixing,
induced at tree-level by $Z^\prime$ exchange as soon as a non-zero
$\Delta_L^{bs}$ coupling is considered. Allowing for a $10\%$
deviation from the SM expectation in the mixing amplitude,
$|M_{12}/M_{12}^{\text{SM}} - 1|<0.1$, one
finds~\cite{Altmannshofer:2014rta}
\begin{equation}
\frac{m_{Z^\prime}}{|\Delta_L^{bs}|} \gtrsim 244 \, \text{TeV} \, .
\end{equation}

\subsection{Lepton flavor constraints}

In principle, the mixings with the vector-like leptons can induce
lepton flavor violating (LFV) processes of the type $\ell_\alpha \to
\ell_\beta \ell_\beta \ell_\gamma$, mediated by the $Z^\prime$
boson. However, one can suppress them with a proper parameter
choice. Since all $b \to s$ anomalies can be simultaneously explained
with (only) new contributions to $\mathcal O_9^\mu$, $\lambda_L^\mu
\neq 0$ is required, but one can choose $\lambda_L^{e,\tau} = 0$. This
would eliminate the coupling of the $Z^\prime$ boson to electrons and
tau leptons, and thus all LFV processes mediated by the $Z^\prime$
boson. This includes LFV in $B$ decays, recently suggested in
\cite{Glashow:2014iga} and further studied in
\cite{Bhattacharya:2014wla,Sahoo:2015wya,Varzielas:2015iva,Boucenna:2015raa}.

Besides the anomalies in $b \to s$ transitions, the LHC might have
found additional hints for new physics. Recently, the CMS
collaboration found a 2.4$\sigma$ excess in the $h\to \tau \mu$
channel which translates into BR$(h\to\tau\mu) = \left(
0.84_{-0.37}^{+0.39} \right)$\%~\cite{Khachatryan:2015kon}. For this
result, the collaboration made use of the 2012 dataset taken at
$\sqrt{s}=8$~TeV with an integrated luminosity of 19.7~fb$^{-1}$. This
large Higgs LFV branching ratio cannot be accommodated in our setup,
since vector-like leptons are known to be unable to reach such LFV
rates in Higgs decays due to limits from the radiative $\tau \to \mu
\gamma$ decay~\cite{Falkowski:2013jya}. However, as suggested in
\cite{Davidson:2010xv,Kopp:2014rva} and recently confirmed in several
works
\cite{Sierra:2014nqa,Heeck:2014qea,Crivellin:2015mga,Dorsner:2015mja},
a simple extension with a second Higgs doublet suffices to explain the
CMS hint.

Finally, we comment on limits from the (former) non-observation of
lepton universality volation. Indeed, very strong bounds have been
derived in many scenarios due to the non-observation of lepton
universality violating effects in, for example, pion and kaon decays
(see for example
\cite{Abada:2012mc,Abada:2013aba,Asaka:2014kia}). However, these are
absent in our model due to the suppression of the vector-like and
first generation SM quark mixing.

\subsection{Precision measurements}

There are several precision measurements in the lepton sector which
might be potentially sensitive to the $Z^\prime$ interactions
considered in this model.

First of all, there is the so-called \emph{neutrino trident
  production}~\cite{Altmannshofer:2014pba}. This is the production of
a muon anti-muon pair by scattering of muon neutrinos in the Coulomb
field of a target nucleus. The cross-section gets additional
contributions in the presence of a new $Z^\prime$ boson that couples
to the muons, and these can be used to constrain $\Delta_L^{\mu \mu}$
and $m_{Z^\prime}$. The impact of this bound on $Z^\prime$ models
designed to solve the $b \to s$ transitions was studied in
\cite{Altmannshofer:2014cfa,Crivellin:2015mga}. Using the CCRF
measurement of the trident cross-section~\cite{Mishra:1991bv}, one can
derive bounds on the mass of the $Z^\prime$ boson and its coupling to
muons. For scenarios with left-handed $Z^\prime$ coupling to muons,
one finds~\cite{Altmannshofer:2014rta}
\begin{equation}
\frac{m_{Z^\prime}}{|\Delta_L^{\mu \mu}|} \gtrsim 470 \, \text{GeV} \, .
\end{equation}

Finally, there are two other constraints related to precision
measurements in leptonic observables: $(g-2)_\mu$ and the $Z$ boson
couplings to the muon (which are modified at the 1-loop level due to
the $Z^\prime$ interactions). However, in \cite{Altmannshofer:2014cfa}
these are found to be relatively mild. For this reason we will not
consider them in our numerical analysis.

\section{Numerical results}
\label{sec:results}

For the numerical analysis we have implemented the model into the {\tt
  Mathematica} package \SARAH
\cite{Staub:2008uz,Staub:2009bi,Staub:2010jh,Staub:2012pb,Staub:2013tta}
and generated {\tt Fortran} code to link this model to \SPheno
\cite{Porod:2003um,Porod:2011nf}. The big advantage of this setup is
that we can cross-check our tree-level approximations with a full
numerical calculation. Moreover, based on the {\tt FlavorKit}
interface \cite{Porod:2014xia}, the generated \SPheno version does not
only calculate the tree-level contributions to the Wilson coefficients
but also all 1-loop corrections. This way, we can see in what
parameter range the tree-level approximation works well and where it
breaks down. In general, one can expect loop effects to become
important when the mixing between SM quarks and the vector-like states
is sizable: this leads to significant changes in the SM-like box and
penguin contributions to $C_9$ and $C_{10}$. This alone would not be
necessarily a problem. However, the same mixing influences also loop
contributions to the other Wilson coefficients, in particular $C_7$
and $C_8$.  These effects have to be small in order to satisfy the
results of global fits to $b \to s$ data.\\ 
As second step, we used
the {\tt CalcHep} \cite{Pukhov:2004ca,Boos:1994xb} output of \SARAH to
implement the model in {\tt MicrOmegas} to calculate the relic
density. Interfacing the parameter values between \SPheno and {\tt
  MicrOmegas} is done via the {\tt SLHA+} functionality of {\tt
  CalcHep} \cite{Belanger:2010st}.

 \begin{figure}[hbt]
 \centering
 \includegraphics[width=0.9\linewidth]{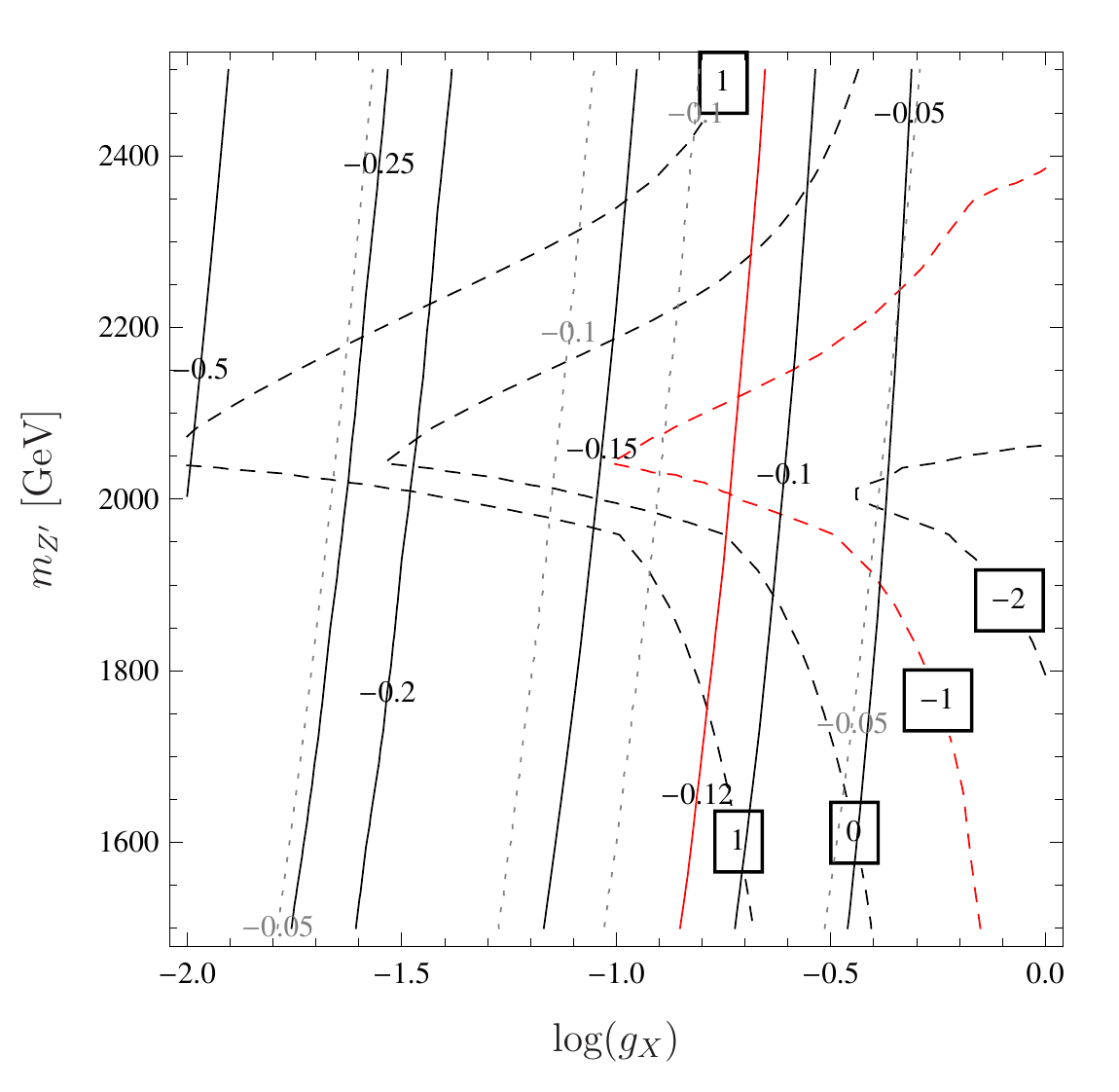}
 \caption{Contours for constant $C_9^\text{NP}/C_9^\text{SM}$ and
   $\log(\Omega_{\text{DM}} h^2)$ (dashed black) in the $(g_X,
   m_{Z^\prime})$ plane. For $C_9^\text{NP}/C_9^\text{SM}$ we show the full
   1-loop results via the black lines, while the dotted grey lines
   give the values using the tree-level approximation only. }
 \label{fig:C9DM}
 \end{figure}

 \begin{figure}[hbt]
 \centering
 \includegraphics[width=0.9\linewidth]{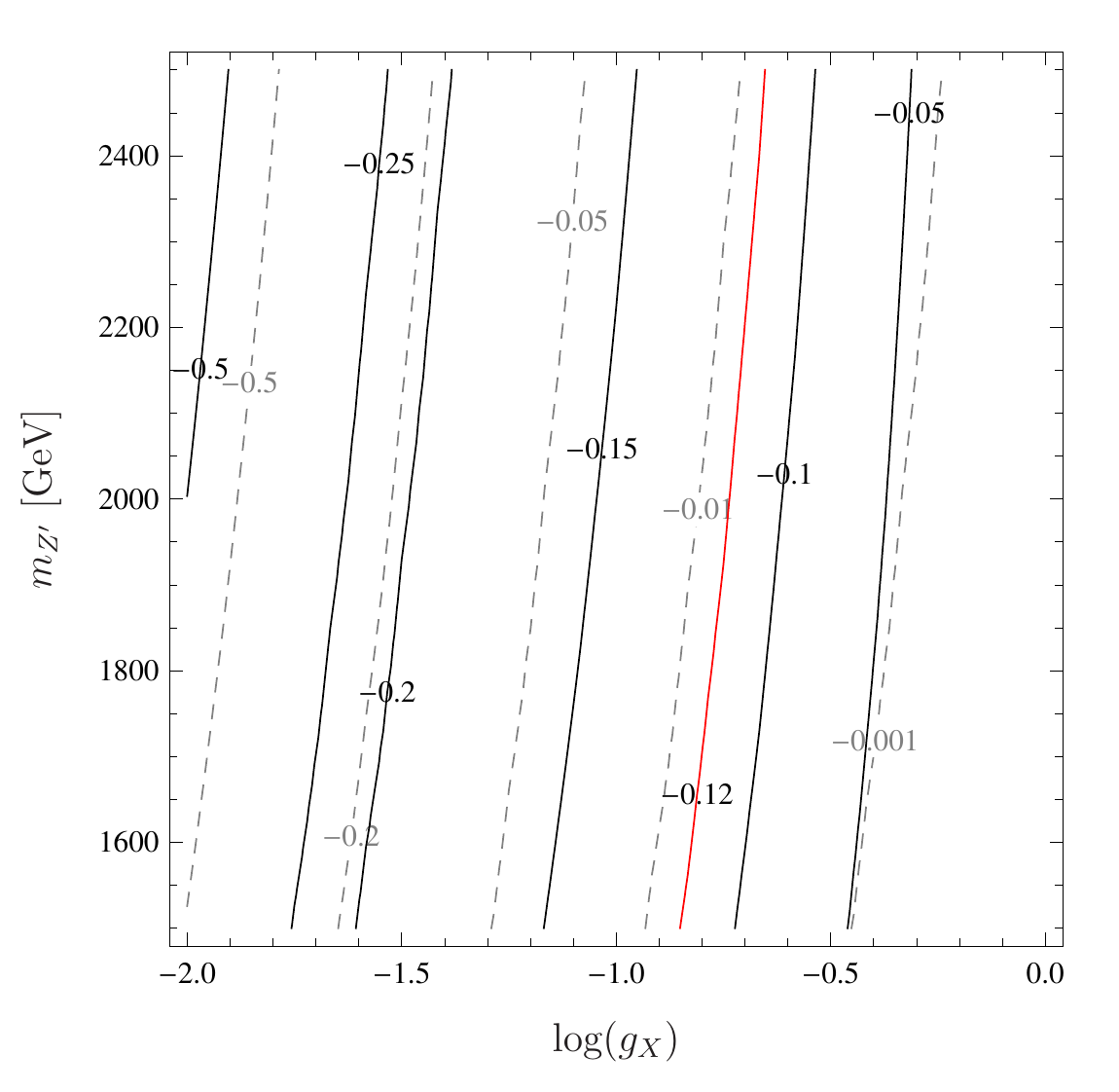}
 \caption{Contours for constant $C_9^\text{NP}/C_9^\text{SM}$ (full black lines)
   and $C_7^\text{NP}/C_7^\text{SM}$ (dotted grey lines) in the $(g_X,
   m_{Z^\prime})$ plane. }
 \label{fig:C7C9}
 \end{figure}

Our main numerical results are summarized in Figs.~\ref{fig:C9DM} and
\ref{fig:C7C9}. In Fig.~\ref{fig:C9DM} we show the dark matter relic density
in the $(g_X, m_{Z^\prime})$ plane, together with the ratio
$C_9^\text{NP}/C_9^\text{SM}$. The other important parameters entering this
calculation have been fixed to
\begin{eqnarray*}
&\lambda_Q^b = \lambda_Q^s=0.025 \,,\quad \lambda_L^\mu = 0.5 &\\ 
& m_Q=m_L=1~\text{TeV} \,,\quad m^2_\chi = 1~\text{TeV}^2 &
\end{eqnarray*}
First of all, we see that with these parameters the model is perfectly
compatible with the constraints discussed in Section
\ref{sec:constraints}. Furthermore, there is a region for moderately
large $g_X \simeq 0.3$ where the DM constraint can be satisfied and
$C^\text{NP}_9 = -12\% \times C^\text{SM}_9$ holds. To reduce the
relic density to $\Omega h^2 \simeq 0.1$ one has to be rather close
the the $Z'$ resonance around 2~TeV.  In the same Figure we show also
$C^\text{NP}_9$ when using the tree-level approximation only. In the
interesting region where the flavor anomalies can be explained, this
approximation is working quite well. However, for decreasing $g_X$ one
would expect smaller values for $C^\text{NP}_9$ at tree-level, while
we find a large increment. The reason is that for constant
$m_{Z^\prime}$ and smaller $g_X$, $v_\phi$ is increasing, leading to a
larger mixing in the quark sector with the vector-like states. Thus,
loop effects become important and quickly dominate the behavior. These
effects are not only expected to show up in $C_9^\text{NP}$ but also
in the other Wilson coefficients as stated above. To demonstrate this,
we plot in Fig.~\ref{fig:C7C9} the contours for
$C^\text{NP}_9/C^\text{SM}_9$ and $C^\text{NP}_7/C^\text{SM}_7$ in the
same plane. Obviously, we find a similar enhancement for small $g_X$
for both coefficients. However, for the most interesting region where
tree-level contributions to $C_9^\text{NP}$ dominates and can explain
the anomalies, the change in $C_7$ is very moderate:
$C^\text{NP}_7/C^\text{SM}_7 < 1\%$.

\section{Summary and discussion}
\label{sec:conclusions}%

We have presented a DM model that successfully addresses the anomalies
in $b \to s$ transitions recently found by the LHCb collaboration. In
our setup, $B$ mesons decay through the $Z^\prime$ into the dark
sector, which mixes with the SM via Yukawa couplings. An interesting
connection between DM and flavor physics emerges, leading to a very
constrained scenario where the NP masses and parameters are restricted
to lie in thin regions of the full parameter space.

The model offers several interesting phenomenological
possibilities. Due to the mixing between the SM muons and the
vector-like lepton doublet, our model predicts a reduction of the
$h-\mu-\mu$ coupling, $g_{h\mu\mu}$, with respect to the SM
prediction. The most direct probe of this coupling is, of course, the
Higgs boson decay into a pair of muons, $h \to \mu^+ \mu^-$,
experimentally very challenging in a hadron collider due to the
expected low branching ratio ($\mathcal{O}(10^{-4})$ in the SM). In
2013, the ATLAS collaboration presented results based on an integrated
luminosity of $20.7$ fb$^{-1}$ in collisions at $\sqrt{s} = 8$ TeV,
without any evidence of a signal~\cite{ATLAS:2013qma}. For $m_h = 125$
GeV, they obtained an upper limit on the $\mu \mu$ signal strength,
$\mu_{\mu \mu} < 9.8$ (at $95\%$ CL). Assuming a SM-like Higgs boson
production, this limit translates into $g_{h\mu\mu} \lesssim 3.1
g_{h\mu\mu}^{\text{SM}}$. Obviously, our model is well within the
limit, since it predicts a reduction of the coupling. Nevertheless, it
is worth pointing out that a precise determination of this coupling in
a future linear collider would be a strong test of the
model~\footnote{According to \cite{Milutinovic-Dumbelovic:2014uta},
  CLIC would be able to measure the $h-\mu-\mu$ coupling with an
  uncertainty of about $19\%$.}.

Let us also comment on the mixings between the vector-like fermions
and the SM ones. These are determined by free Yukawa parameters which,
in principle, are naturally expected to be of the same order for all
SM generations. However, in order to avoid conflict with existing
experimental data, one is forced to strongly suppress the mixings with
the first generation, implying $\lambda_Q^{d},\lambda_L^{e} \ll
\lambda_Q^{s,b},\lambda_L^{\mu,\tau}$. This might seems like an ad-hoc
assumption. However, it just reflects the \emph{constant} need for a
theory of flavor, also required to understand the Yukawa structure of
the SM~\footnote{We note that although the lepton sector of
  \cite{Altmannshofer:2014cfa} naturally explains the required
  hierarchy for the lepton couplings, in the quark sector the
  situation is exactly the same as in our model.}.

Finally, one can envisage several phenomenological directions in which
this model can be further explored. By allowing for non-zero couplings
to first generation quarks and leptons, the impact of the model gets
extended to many additional observables of interest. In particular,
the LHC would be able to probe the parameter space of the model via
the production of the $Z^\prime$ boson. Furthermore, since the scalar
DM carries, at early times, a conserved $U(1)_X$ charge it will
develop an asymmetry, provided chemical equilibrium with the thermal
bath is guaranteed. In such a case (asymmetric DM scenario
\cite{Boucenna:2013wba}), a link between the baryon and DM asymmetries
is something worth exploring.

\section*{Note added}
An update of the global fit \cite{Altmannshofer:2014rta} was recently
presented in \cite{Altmannshofer:2015sma}.  While this would change
slightly the numerical output of our analysis, our conclusions would
remain unchanged.

\section*{Acknowledgments}
We are grateful to Javier Virto and Jorge Martin Camalich for
enlightening discussions. DAS wants to thank Diego Restrepo and Carlos
Yaguna for very useful conversations on dark matter. DAS is supported
by a ``Charg\`e de Recherches'' contract funded by the Belgian FNRS
agency. AV acknowledges partial support from the
EXPL/FIS-NUC/0460/2013 project financed by the Portuguese FCT.

\end{document}